\documentclass{article}

\usepackage{ifthen}

\begin{document}

\title{Influence Of Current Leads On Critical Current For Spin Precession In Magnetic Multilayers}

\author{L. BERGER}

\date{Physics Department, Carnegie Mellon University, Pittsburgh, PA 15213}

\maketitle

\setlength{\baselineskip}{4mm}

In magnetic multilayers, a dc current induces a spin precession above a certain critical current. Drive torques responsible for this can be calculated from the spin accumulation $\overline{\Delta\mu}$ . Existing calculations of $\overline{\Delta\mu}$ assume a uniform cross section of conductors. But most multilayer samples are pillars with current leads flaring out immediately to a much wider cross-section area than that of the pillar itself. We write spin-diffusion equations of a form valid for variable cross section, and solve the case of flat electrodes with radial current distribution perpendicular to the axis of the pillar. Because of the increased volume available for conduction-electron spin relaxation in such leads, $\overline{\Delta\mu}$ is reduced in the pillar by at least a factor of 2 below its value  for uniform cross section, for given current density in the pillar. Also, $\overline{\Delta\mu}$ and the critical current density for spin precession become nearly independent of the thickness of the pinned magnetic layer, and more dependent on the thickness of the spacer, in better agreement with measurements by Albert et al. (2002).

\vspace{1ex}


L. Berger, Physics Dept., Carnegie Mellon University, Pittsburgh, PA 15213;
Tel.: 412 268 2919; Fax: 412 681 0648; E-mail: lberger@andrew.cmu.edu

\vspace{1ex}

\hspace{6em} INTRODUCTION

\vspace{1ex}

Through the s-d interaction, a dc electrical current can cause $^{1,2}$ an instability of localized magnetic spins $\bf{S}_{2}$ in the free magnetic layer $F_{2}$ of a multilayer (Fig. 1a) containing of two magnetic layers $F_{1},F_{2}$. The purpose of the pinned magnetic layer $F_{1}$ is to prepare the conduction-electron spins in a fixed direction parallel or antiparallel to its magnetic spins $\bf{S}_{1}$, before they enter $F_{2}$ through the nonmagnetic spacer $N_{2}$ to generate  drive torques on $\bf{S}_{2}$.

The Slonczewski theory $^{1}$ based on spin current and the Berger theory $^{2}$ based on spin accumulation $\Delta\mu$ can be shown to be more or less equivalent. In their original form, they were both associated with a translation of the Fermi surface in momentum space, caused by the current (Fig. 1b). Near an interface, there is also $^{3}$ a current-induced isotropic expansion of the Fermi surface of one spin, and contraction of the other (Fig. 1c). This is connected with the spin accumulation of Silsbee and Johnson $^{4}$ which we denote by $\overline{\Delta\mu}$, and which is linked to the current density only in a nonlocal manner. A 2002 paper by Slonczewski $^{1}$ uses different spin accumulations on the two halves of the Fermi surface, a scheme equivalent to our $\Delta\mu$ and $\overline{\Delta\mu}$.

We will use scalar spin accumulations $\Delta \mu$ and $\overline{\Delta\mu}$ by opposition to vectorial ones, since we consider current densities not exceeding the critical density where all spins are still nearly parallel to each other.

In most earlier detailed calculations of  $\overline{\Delta\mu}$ in multilayers $^{3,5-7}$, the cross section of current leads was assumed constant and the same as in the multilayer itself (Fig 1a). In that one-dimensional problem, $\overline{\Delta\mu}$ was found $^{3,7}$ to be about 10 times larger than the spin accumulation $\Delta\mu$ arising from translation, and therefore dominant.

The purpose of the present paper is to treat the more realistic case where current leads flare out to much larger cross-section area. This results in a much improved agreement between theory and experimental results.

\vspace{1ex}

\hspace{3em} SPIN DIFFUSION WITH VARIABLE CROSS SECTION AREA

\vspace{1ex}

We introduce a coordinate r along the length of a conductor. The current densities $\bf{j}_{\uparrow},\bf{j}_{\downarrow}$ for spin up and spin down are assumed to have constant magnitudes over a conductor cross-section surface corresponding to r = const., of area A(r), and to be normal to that surface. Electron-spin relaxation gives $dI_{\uparrow}/dr=-dI_{\downarrow}/dr=egA(r)\overline{\Delta\mu}/2\tau_{sr}$ where $I_{\uparrow},I_{\downarrow}$ are the spin-up and spin-down currents, and g the electron density of states per unit volume at the Fermi level. Also, $\overline{\Delta\mu}=\overline{\mu}_{\uparrow}-\overline{\mu}_{\downarrow}$ is the spin accumulation coming from expansion/contraction, and $\overline{\mu}_{\uparrow},\overline{\mu}_{\downarrow}$ the spin-up and spin-down electrochemical potentials. Finally, $\tau_{sr}$ is the conduction-electron spin-relaxation time. The combination of Ohm's and Fick's laws reads $I_{\uparrow}=(\sigma_{\uparrow}A(x)/e)d\overline{\mu}_{\uparrow}/dr;I_{\downarrow}=(\sigma_{\downarrow}A(x)/e)d\overline{\mu}_{\downarrow}/dr$, where $\sigma_{\uparrow}=\rho_{\uparrow}^{-1},\sigma_{\downarrow}=\rho_{\downarrow}^{-1}$ are the spin-up and spin-down conductivities. By introducing $\Delta I=I_{\uparrow}-I_{\downarrow}$, the equations above can be rewritten in a more convenient form: 
\begin{eqnarray}
\frac{d(\overline{\Delta\mu}/K)}{dX}=\frac{\Delta I-\Delta I_{0}}{A(r)};\frac{d\Delta I}{dX}=A(r)\frac{\overline{\Delta\mu}}{K};X=r/l_{sr};\\
\Delta I_{0}=I\frac{\rho_{\downarrow}-\rho_{\uparrow}}{\rho_{\uparrow}+\rho_{\downarrow}};K=el_{sr}(\rho_{\uparrow}+\rho_{\downarrow})/2;l_{sr}=(\frac{2\tau_{sr}}{e^{2}g(\rho_{\uparrow}+\rho_{\downarrow})})^{1/2},
\end{eqnarray}  

where $l_{sr}$ is the local spin-diffusion length and $I=I_{\uparrow}+I_{\downarrow}$ the total current. In the special case A(r)= const., Eqs. (1-2) reduce to Eqs. (1-2) of Ref. 7, where a constant cross section of conductors was assumed.

Many of published experimental results $^{8}$ about current-induced spin precession have been obtained with multilayer samples consisting of a pillar of uniform radius R, with current leads in the shape of flat electrodes where the current runs in radial directions normal to the pillar axis (Fig. 2a). In such electrodes, we use as the coordinate r the radial distance of a point P from the pillar axis.

To simplify calculations, we assume that the direction of the current in the electrodes remains radial even near the pillar, at r values as low as r=R. Then, the cross-section surfaces are cylinders (Fig. 2a) centered on the pillar axis, of area $A(r)=2\pi rL_{e}=2\pi l_{sr}L_{e}X$, where $L_{e}$ is the thickness of the flat electrode. We combine Eqs. (1) and (2) after defining $\Delta i(X)=\Delta I/2\pi RL_{e}$, and obtain the differential equation $d^{2}(\Delta i)/dX^{2}-X^{-1}d(\Delta i)/dX-\Delta i=0$, which we solve by Mathematica for $\Delta i(X)$, with the boundary condition $\Delta i(\infty)=0$ and the normalization condition $\Delta i(X=R/l_{sr})=1$. Since $\Delta I_{0}=0$ in non-magnetic leads, the Eqs. (1-2) are homogeneous , and the normalization of $\Delta i(X)$ is arbitrary. Then we obtain $\overline{\Delta\mu}(X)/K_{N}$ by the second Eq. (1). Here, $K_{N}$ is the value of K in the Cu electrodes. We plot $\Delta i(X)$ and $-\overline{\Delta\mu}(X)/K_{N}$ in Fig. 2b for $X\geq R/l_{sr}$.

The location r=R in the electrode near the pillar (Fig. 2a) corresponds to $X=R/l_{sr}=0.433$, assuming a reasonable value R=65 nm. We use $l_{sr}=150 nm$ for Cu, close to a value 140 nm measured $^{9}$ at 77 K. Also $^{10}$, $\rho =3\times 10^{-8}\Omega m$, leading to $K_{N}=1.44\times 10^{-33}m^{2}J/A$. For our purpose, the most important quantity is the ratio $(\overline{\Delta\mu}(X=0.433)/K_{N})/\Delta i(X=0.433)$ at that location, equal (Fig. 2b) to -0.527. Assuming a reasonable $^{8}$ electrode thickness $L_{e}=100 nm$, and introducing $\Delta j=j_{\uparrow}-j_{\downarrow}$, where $j_{\uparrow},j_{\downarrow}$ are the spin-dependent current densities in the pillar, assumed parallel to the pillar axis, current conservation gives $\Delta j=-3.08\Delta i(X=0.433)=-3.08$. Thus $(\overline{\Delta\mu}(X=0.433)/K_{N})/\Delta j=0.171$. This compares to the value $(\overline{\Delta\mu}/K_{N})/\Delta j=1$ of the ratio which holds $^{7}$ in leads that have the same constant  cross section as the multilayer itself.  On the other hand, the electrodes still do not act like a complete ``spin short circuit'' with a ratio equal to zero.

The depressed value of this ratio can be attributed to the large volumes of copper available nearby in the wide and thick electrodes, for relaxation of the spin of conduction electrons. As we will see later, this will also lead to a depressed value of $\overline{\Delta\mu}$ itself in the multilayer.

\vspace{1ex}

\hspace{2em} SPIN DIFFUSION IN LAYERS $N_{1}$ AND $N_{3}$

\vspace{1ex}

For simplicity, the current direction in nonmagnetic layers $N_{1}$ and $N_{3}$ (Fig. 2a) is assumed to be everywhere parallel to the pillar axis. Also, these layers are assumed to be made of copper, with thicknesses $L_{N1}=L_{N3}=5nm$. Using Eqs. (1-2), the variations of $\overline{\Delta\mu}/K_{N}$ and of $\Delta j$ across the thickness of $N_{1}$ are found to be $\Delta jL_{N1}/l_{sr}\simeq -0.102$ and $(\overline{\Delta\mu}/K_{N})L_{N1}/l_{sr}\simeq -0.017$, respectively. Starting from the values of $\overline{\Delta\mu}/K_{N}$ and $\Delta j$ near the electrode at r=R given in the last section and applying these corrections, we obtain $\overline{\Delta\mu}/K_{N}=-0.629$ and $\Delta j=-3.10$ in $N_{1}$ at the $N_{1}/F_{1}$ interface (Fig. 2a). This results in a ratio $(\overline{\Delta\mu}/K_{N})/\Delta j=0.20$ at that location. We conclude that the presence of layer $N_{1}$ causes only a minor increase of the value of the ratio.

The presence of the transition region ABCD connecting the electrode to the pillar probably has also a minor effect on the ratio, which we ignore as it is more difficult to calculate.

The problem of the right-hand side electrode and of $N_{3}$ (Fig.2a) can be treated similarly, leading to a ratio $(\overline{\Delta\mu}/K_{N})/\Delta j=-0.20$ at the $F_{2}/N_{3}$ interface in $N_{3}$. This compares $^{7}$ to a value -1 for a lead of constant cross section.  

\vspace{1ex}

\hspace{1em} SPIN DIFFUSION IN MULTILAYER

\vspace{1ex}

Using Mathematica, we solve Eqs. (1-2) in layers $F_{1}$ and $F_{2}$ made of Co, with constant cross-section area A. Here, K has $^{7}$ the value $K_{F}=3.4\times 10^{-33} m^{2}J/A$. Also, $\Delta I_{0}/I=0.36$. As in Ref. 7, these are derived from values $\rho_{\uparrow}=23\times 10^{-8}\Omega m, \rho_{\downarrow}=49\times 10^{-8}\Omega m$ and spin-diffusion length $l^{F}_{sr}=59nm$ measured for Co nanowires by Piraux et al. $^{10}$ at 77 K. Although the critical-current data we will try to explain are for room temperature, we use these low-temperature parameter values in order to partly compensate for the larger electron-scattering rate observed in nanowires as compared to thin films. This also applies to parameter values quoted for Cu in earlier sections. Another reason to use these values is to be able to compare our results to those of Ref. 7. Instead of r, the coordinate in the pillar is called x, the origin being at the $N_{2}/F_{2}$ interface. We are not free anymore to arbitrarily normalize $\overline{\Delta\mu}$ and $\Delta j$ as was done in previous sections. We calculate $\overline{\Delta\mu}/K_{F}$ in the same way as in Ref. 7, except that the variation across spacer $N_{2}$ cannot be neglected now. Discontinuous variations of $\overline{\Delta\mu}/K_{F}$ and $\Delta I$ at $N_{1}/F_{1}, F_{1}/N_{2}, N_{2}/F_{2}$ and $F_{2}/N_{3}$ interfaces are given $^{7}$ by  

\begin{eqnarray}
D(\overline{\Delta\mu}/K_{F})=\frac{K_{s}}{K_{F}}\frac{\Delta I-\Delta I_{s}}{A};D(\Delta I/A)=\frac{r^{*}}{\rho^{*}l^{F}_{sr}}\overline{\Delta\mu}/K_{F};\\
\Delta I_{s}=I\frac{r_{\downarrow}-r_{\uparrow}}{r_{\downarrow}+r_{\uparrow}};K_{s}=e(r_{\uparrow}+r_{\downarrow})/2, \nonumber
\end{eqnarray}

where $^{7,10}$ $\Delta I_{s}/I=0.85,K_{s}=9.6\times 10^{-35} m^{2}J/A,l^{F}_{sr}=59nm,r^{*}=3\times 10^{-16}\Omega m^{2},\rho^{*}=18\times 10^{-8}\Omega m$ at low temperature. Here, $r^{*},r_{\uparrow},r_{\downarrow}$ and $\rho^{*}$ characterize the electrical resistance of a Co/Cu interface and of a Co layer, respectively.

The variations of $\overline{\Delta\mu}/K_{F}$ and $\Delta j=\Delta I/A$ across spacer $N_{2}$ are calculated in the same way as explained for $N_{1}$ and $N_{3}$ in the last section. Also, as in Eqs. (7-8) of Ref. 7, the current leads, including layers $N_{1}$ and $N_{3}$, influence the boundary conditions at the $N_{1}/F_{1}$ interface through the ratio discussed in the last section:

\begin{equation}
\frac{\overline{\Delta\mu}/K_{N}}{\Delta I/A}=0.20,
\end{equation}

where the quantities are evaluated in $N_{1}$. Similarly in $N_{3}$ at the $F_{2}/N_{3}$ interface, the boundary condition reads:

\begin{equation}
\frac{\overline{\Delta\mu}/K_{N}}{\Delta I/A}=-0.20.
\end{equation}

The results of the calculation for $\overline{\Delta\mu}/K_{F}$ are shown versus $x/l^{F}_{sr}$ in Fig. 3, assuming reasonable thicknesses $L_{N1}=5nm, L_{F1}=10nm, L_{N2}=6nm,L_{F2}=2.5nm,L_{N3}=5nm$, and a value of $1A/m^{2}$ for the current density $j=I/A$ in the pillar. We consider the parallel state (P) where $F_{1}$ and $F_{2}$ are magnetized in the same direction. The present case of flat electrodes with radial current distribution is labelled ``variable cross section''. For comparison, the case $^{7}$ of current leads of cross section equal to that of the multilayer is labelled ``constant cross section''.

In the free layer $F_{2}$ near the $N_{2}/F_{2}$ interface active in spin transfer, i.e., at $x/l^{F}_{sr}=0$, we find (Fig. 3) $\overline{\Delta\mu}/K_{F}=-0.01312A/m^{2}$ and $\Delta I/A=0.338A/m^{2}$ in the case of variable cross section. This compares $^{7}$ to $\overline{\Delta\mu}/K_{F}=-0.0333A/m^2$ and $\Delta I/A=0.1505A/m^{2}$ with constant cross section. We see that the effect of the flat electrodes is to decrease $|\overline{\Delta\mu}/K_{F}|$ and increase $\Delta I/A$. Also, Fig. 3 shows that $\overline{\Delta\mu}/K_{F}$ varies faster with $x/l^{F}_{sr}$ than before in nonmagnetic layers, and much slower in magnetic layers. This results from the first Eq. (1) and from $\Delta I/I$ now being nearly equal to the value of $\Delta I_{0}/I=0.36$ which holds in magnetic layers. Only in the electrodes does $\Delta I/I$ drop much below that value. 

We have also calculated $\overline{\Delta\mu}/K_{F}$ in $F_{2}$ at $x/l^{F}_{sr}=0$ as a function of thickness $L_{F1}$ (Fig. 4, curve labelled ``P+variable cross section''), for the same values of $L_{N1},L_{N2},L_{F2},L_{N3}$ as above, and $I/A=1A/m^{2}$. The fact that $\overline{\Delta\mu}/K_{F}$ is nearly independent of $L_{F1}$ is directly related to it being nearly independent of $x/l^{F}_{sr}$ in $F_{1}$ (Fig. 3). In the case of leads with constant cross section (Fig. 4, curve labelled ``P+constant cross section), $\overline{\Delta\mu}/K_{F}$ increased with increasing $L_{F1}$.

Using the same methods, we have calculated $\overline{\Delta\mu}/K_{F}$ and $\Delta I/A$ at $x/l^{F}_{sr}=0$ in $F_{2}$  for the antiparallel (AP) state where the magnetization of $F_{2}$ has been reversed (Fig. 4, curve labelled ``AP+variable cross section''). Now, $\Delta I_{0}/I=-0.36$ in $F_{2}$ and $\Delta I_{s}/I=-0.85$ at the $N_{2}/F_{2}$ and $F_{2}/N_{3}$ interfaces. The effect of leads of variable cross section is much less pronounced in the AP than in the P state. For example, $\Delta I/A$ is now only 0.0833 in the multilayer when $L_{F1}=10nm$, not much more than its value $^{7}$ of 0.0384 with leads of constant cross section. Correspondingly, $|\overline{\Delta\mu}/K_{F}|$ is now almost as large (Fig 4) as it was with leads of constant cross section, and also increases with increasing $L_{F1}$.

\vspace{1ex}

\hspace{6em}  CRITICAL CURRENT

\vspace{1ex}

As mentioned in the Introduction, the spin accumulation $\Delta\mu$ in the pillar arising from Fermi-surface translation is almost negligible $^{7}$ as compared to the accumulation from expansion/contraction, in the case of leads of constant cross section. However, as shown in the last section, $|\overline{\Delta\mu}|$ is smaller and $\Delta I/A$ larger in the case of leads of variable cross section. Since $^{2}$ $\Delta\mu\simeq -(\hbar k_{N}/en^{N}_{e})\Delta I/A$, where $k_{N}$ and $n^{N}_{e}$ are the electron wavenumber and electron density in the spacer $N_{2}$, $\Delta\mu$ may become comparable to $\overline{\Delta\mu}$ in that case. We calculate $\Delta\mu$ from the formula above, with $k_{N}=1.36\times 10^{10}m^{-1},n^{N}_{e}=8.5\times 10^{28}m^{-3}$ as in Cu, and find indeed $\overline{\Delta\mu}/\Delta\mu=1.26$ for the P state  in $F_{2}$ at x=0, assuming $L_{F1}=10 nm$. Note, however, that $\Delta\mu$ is much less important for the AP state, with $\overline{\Delta\mu}/\Delta\mu=19.1$.

The critical current density for incipient spin precession is $^{11}$

\begin{equation}
j_{c}=\frac{\pm \hbar \omega_{eff}}{(\overline{\Delta\mu}+\Delta\mu)/(I/A)}(1+\frac{\alpha_{b}}{\alpha_{s}}).
\end{equation}

where $\omega_{eff}=\gamma (M_{s}/2+H_{z})$ in SI units, $\alpha_{b}$ is the bulk Gilbert damping parameter, and $\alpha_{s}$ the surface Gilbert parameter at the $N_{2}/F_{2}$ interface. It is through $\alpha_{s}$ that the spin-transfer properties of the interface are taken into account. The - and + signs apply to the P and AP states, respectively. The field $H_{z}$ is in the plane of layers, but is assumed zero for the time being.

From Eq. (6), we obtain the ratio of critical current densities for the P and AP states:

\begin{equation}
\frac{|j^{P}_{c}|}{|j^{AP}_{c}|}=\frac{(\overline{\Delta\mu}+\Delta\mu)_{AP}}{(\overline{\Delta\mu}+\Delta\mu)_{P}}.
\end{equation}

Using Eq. (7), we calculate $|j^{P}_{c}|/|j^{AP}_{c}|$ from the $\Delta\mu$ values above and from the $\overline{\Delta\mu}$ values obtained for $L_{F2}=2.5nm$ in the last section, and plot it versus $L_{F1}$. The results are shown as the curves labelled ``spin diffusion'' in Fig. 5a, for leads of variable cross section. The circles indicate experimental $|j^{P}_{c}|/|j^{AP}_{c}|$ values from various authors $^{8}$, for pillar samples similar to that of Fig. 2a.

Even when the applied field is zero, fixed magnetic layer $F_{1}$ produces (Fig. 5b) a dipolar field $H^{D}_{z}<0$ in $F_{2}$ at the important $N_{2}/F_{2}$ interface. Assuming a magnetically saturated $F_{1}$, it is in SI units $H^{D}_{z}=-R^{2}L_{F1}M_{s}/4d^{3}$, where $d\simeq (R^{2}+(L_{F1}/2+L_{N2})^{2})^{1/2}$ is the average distance between magnetic poles and a point of the interface. Then, using Eq.(6) and values R = 65 nm and $\mu_{0}M_{s}=1.8\ T$ for Co, the ratio $|j^{P}_{c}|/|j^{AP}_{c}|$ obtained from Eq. (7) must be corrected by a factor $(M_{s}/2+H_{K}+H^{D}_{z})/(M_{s}/2+H_{K}-H^{D}_{z})<1$. The results are shown as the curves labelled ``spin diffusion+dipolar'' in Fig. 5a, assuming a reasonable value $\mu_{0}H_{K}=0.1\ T$ of the in-plane anisotropy field. The effect of the dipolar field is in all cases to reduce the value of $|j^{P}_{c}|/|j^{AP}_{c}|$ by an amount increasing with increasing $L_{F1}$, giving a better fit to the experimental values. Note that the actual samples $^{8}$ do not all have the radius R = 65 nm or the exact shape assumed here. Also, the above formula for d tends to overestimate the average d value at points of the $N_{2}/F_{2}$ interface not on the pillar axis.

From Eq. 6 and from the $\overline{\Delta\mu}$ and $\Delta\mu$ obtained above for the P and AP states, we derive normalized values of the critical current densities $j^{P}_{c}$ and $j^{AP}_{c}$ for leads of variable cross section, which are plotted versus $L_{F1}$ in Fig. 6. In addition, we plot the normalized value of $j^{P}_{c}-j^{AP}_{c}$.

Albert et al. $^{12}$ have measured $j^{P}_{c}-j^{AP}_{c}$ for leads of variable cross section similar to those considered here. The best fit of their data (for $L_{F2}=2.5nm$) to the corresponding theoretical curve in Fig. 6 is obtained for $\alpha_{b}/\alpha_{s}=5.04$, and is shown as the dashed line. To minimize the effect of non-zero temperature, we use data for fast current rise. In agreement with the experimental values, the theoretical values of $j^{P}_{c}-j^{AP}_{c}$ for leads of variable cross section in Fig. 6 are nearly independent of $L_{F1}$. If we had assumed constant cross section, the theoretical value of that quantity would have considerably decreased with increasing $L_{F1}$, giving much less agreement.

 In turn, using the rough value $\alpha_{s}\geq 0.0073$ predicted for Co by Eq. (20) of the first Ref. 2, the above $\alpha_{b}/\alpha_{s}$ leads to $\alpha_{b}\geq 0.0368$, considerably more than the value 0.007 obtained $^{13}$ from ferromagnetic resonance in very thin Co films. The origin of this discrepancy is not clear. Maybe the value of $\alpha_{s}$ ,and of the corresponding current-induced drive torques, is lower than predicted.

We have also calculated $\overline{\Delta\mu}/K_{F}$ and $\Delta I/A$ versus thickness $L_{N2}$ of spacer $N_{2}$, assuming variable lead cross section. We use fixed thicknesses $L_{N1}=L_{N3}=5nm, L_{F1}=10nm, L_{F2}=2.5nm$. From these $\Delta I/A$ we obtain $\Delta\mu$ as before. Then, using Eq. (6), normalized values of the critical current density $j^{P}_{c}$ for the P state are given by
 
\begin{equation}
\frac{j^{P}_{c}(L_{N2})}{j^{P}_{c}(L_{N2}=6nm)}=\frac{(\overline{\Delta\mu}+\Delta\mu)_{L_{N2}=6nm}}{(\overline{\Delta\mu}+\Delta\mu)_{L_{N2}}},
\end{equation}

and similarly for the AP state. The results are plotted versus $L_{N2}$, and appear as two dashed curves in Fig. 7. We see that both $j^{P}_{c}$ and $j^{AP}_{c}$ increase with increasing $L_{N2}$, the increase of $j^{P}_{c}$ being larger by a factor $\simeq 10$. Albert et al.$^{12}$ have measured the critical current densities versus $L_{N2}$ (though with $L_{F1}=8nm,L_{F2}=2nm$), and their values are shown as solid and open squares for the P and AP states, respectively. Like us, they find that $j^{P}_{c}$ and $j^{AP}_{c}$ increase, and that $j^{P}_{c}$ increases faster. The increases are in a ratio of $\simeq 3$. In our theory, these variations of $j^{P}_{c}$ and $j^{AP}_{c}$ with $L_{N2}$ reflect directly the spatial variation of $\overline{\Delta\mu}/K_{F}$ across the spacer $N_{2}$ in Fig. 3. It is therefore more pronounced for variable than for constant lead cross section, leading to better agreement with experiments.

Albert et al. $^{12}$ have explained quantitatively these experimental results with a model which, for $j^{P}_{c}$, involves only spin-down electrons reflected at the $F_{1}/N_{2}$ interface. However, the spin-down reflection coefficient at a Co/Cu interface being $^{6}$ 0.39, only a fraction of incident spin-down electrons would take part in such a process. Current-dependent torques by spin-down and spin-up electrons transmitted from $F_{1}$ to $F_{2}$ should also be considered if the model is to be quantitative. We have been able to explain qualitatively these same results (Fig. 7) with a theory where Fig. 3 shows that interface reflections and scattering inside the layers cause comparable spatial variations of the spin accumulation.

Finally, the dependence of current-induced drive torques on the thickness $L_{F2}$ of the free magnetic layer was already calculated in our 1997 paper $^{2}$, for the Fermi-surface translation mechanism. We found the torque to be proportional to $L_{F2}^{-1}$, with added oscillations periodic in $L_{F2}$. Except for the oscillations, this is also consistent with the Albert data $^{12}$.

\vspace{1ex}

\hspace{4em} RELATION TO GIANT MAGNETORESISTANCE

\vspace{1ex}

It has been shown $^{14}$ that the giant magnetoresistance $\Delta R$ of a magnetic multilayer  may be increased by inserting a thin layer of a material with fast electron-spin relaxation near the current leads. This is just the opposite of the situation for the spin accumulation $\overline{\Delta \mu}$ as described in the present paper, where $|\overline{\Delta \mu}|$ is maximized by reducing the total rate of spin relaxation in the leads. 

By using an equivalent dc electrical circuit which describes $^{15}$ both conduction and spin relaxation in the sample, it is easy to confirm that $\Delta R$ and $|\overline{\Delta \mu}|$ are influenced very differently by spin relaxation in the leads.

\vspace{1ex}

\hspace{2em} MINIMIZATION OF CRITICAL CURRENT

\vspace{1ex}

For technical applications, it is important to minimize the critical current densities $j^{P}_{c},j^{AP}_{c}$. According to Eq. 6, one strategy would be to maximize $|\overline{\Delta\mu}|$ for given current density. As we saw earlier, this would require increasing the thickness of nonmagnetic layers $N_{1},N_{3}$ (Fig. 2a), to values approaching the spin diffusion length. This would move the electrodes further away from the important magnetic layers $F_{1},F_{2}$, thus increasing the ratio of $\simeq 0.2$ appearing in Eq. 4 and increasing $|\overline{\Delta\mu}|$ itself. After that is done, increasing the thickness of fixed magnetic layer $F_{1}$ would also be useful. A second way to maximize $|\overline{\Delta\mu}|$ and, in turn, reduce the critical current density might be to increase the constant $K_{N}$ in Eq. 4 by  making the electrodes and layers $N_{1},N_{3}$ of a light material with slow spin relaxation, such as aluminum.

As we saw in the section on critical current, $\Delta\mu$ is sometimes comparable to $\overline{\Delta\mu}$ in the case of leads with variable cross section. This suggests an opposite strategy of maximizing $|\Delta\mu|$ instead of $|\overline{\Delta\mu}|$, in Eq. 6. Since $\Delta\mu\propto \Delta I$, this means that we should decrease the ratio of 0.2 appearing in Eq. 4 by reducing or eliminating layers $N_{1},N_{3}$; or that we should increase the constant $K_{N}$ in Eq. 4 by making these layers and the electrodes of a spin-relaxing material. Indeed, Urazhdin et al. $^{16}$ have caused a decrease of both $j^{P}_{c}$ and $j^{AP}_{c}$ by a factor of $\simeq 2$ through insertion of a relaxing $Fe_{50}Mn_{50}$ layer near one lead.

Which one of these two strategies works best may depend on whether $\overline{\Delta\mu}$ or $\Delta\mu$ happens to be initially larger, which in turn depends on the initial lead configuration. Note, however, that Eqs. 1 show that $\Delta I/A$ can never exceed $\Delta I_{0}/A\simeq 0.36$, so that the $|\Delta\mu|$ value itself is limited. 

\vspace{1ex}

\hspace{2em} CONCLUSIONS AND FINAL REMARKS

\vspace{1ex}

Most pillar samples used in experiments of current-driven spin precession in magnetic multilayers have current leads of cross section much larger than that of the pillar itself. By treating a specific geometry, we show that one kind $\overline{\Delta\mu}$ of spin accumulation responsible for these oscillations is reduced significantly by the presence of such leads, and the other $\Delta\mu$ is increased. Also, the critical current density for spin precession becomes less dependent on the thickness of the fixed magnetic layer $F_{1}$ and more dependent on the thickness of the spacer $N_{2}$, bringing better agreement with existing experimental results by Albert et al. $^{12}$. One limitation is that we have assumed that the two current leads have exactly the same geometry and composition, which is not quite realistic $^{8}$.

While the original 1996 theories of spin transfer $^{1,2}$ assumed smooth interfaces which conserve the momentum component parallel to the interface, or assumed a specific model of the Fermi surface, the present work does not need such assumptions.

I am grateful to A. Rebei, J. Zhu and Ya.B. Bazaliy for useful discussions.

\vspace{1ex}

\hspace{6em} REFERENCES

\vspace{1ex}

1. J.C. Slonczewski, J. Magn. Magn. Mater. $\underline{159}$, L1 (1996); $\underline{195}$, L261 (1999); $\underline{247}$, 324 (2002).

2. L. Berger, Phys. Rev. B $\underline{54}$, 9353 (1996); J. Appl. Phys. $\underline{81}$, 4880 (1997). 

3. L. Berger, IEEE Trans. Magn. $\underline{34}$, 3837 (1998); J. Appl. Phys. $\underline{89}$, 5521 (2001).

4. M. Johnson and R.H. Silsbee, Phys. Rev. Lett. $\underline{55}$, 1790 (1985).

5. T. Valet and A. Fert, Phys. Rev. B $\underline{48}$, 7099 (1993); A. Fert and S.F. Lee, Phys. Rev. B, $\underline{53}$, 6554 (1996).

6. M.D. Stiles and A. Zangwill, J. Appl. Phys. $\underline{91}$, 6812 (2002).

7. L. Berger, J. Appl. Phys. $\underline{93}$, 7693 (2003).

8. J.A. Katine, F.J. Albert, R.A. Buhrman, E.B. Myers and D.C. Ralph, Phys. Rev. Lett. $\underline{84}$, 3149 (2000); F.J. Albert, J.A. Katine, R.A. Buhrman and D.C. Ralph, Appl. Phys. Lett. $\underline{77}$, 3809 (2000); J. Grollier, V. Cros, H. Hamzic, J.M. George, H. Jaffres and A. Fert, Appl. Phys. Lett. $\underline{78}$, 3663 (2001); E.B. Myers, F.J. Albert, J.C. Sankey, E. Bonet, R.A. Buhrman and D.C. Ralph, Phys. Rev. Lett. $\underline{89}$, 196801 (2002); F.J. Albert, N.C. Emley, E.B. Myers, D.C. Ralph and R.A. Buhrman, Phys. Rev. Lett. $\underline{89}$, 226802 (2002); J.Z. Sun, D.J. Monsma, T.S. Kuan, M.J. Rooks, D.W. Abrahams, B. Oezyilmaz, A.D. Kent and R.H. Koch, J. Appl. Phys. $\underline{93}$, 6859 (2003).

9. L. Piraux, S. Dubois, A. Fert, J. Magn. Magn. Mater. $\underline{159}$, L287 (1996).

10. L. Piraux, S. Dubois, A. Fert, and L. Belliard, Eur. Phys. J. B \underline{4}, 413 (1998).

11. L. Berger, J. Appl. Phys. $\underline{91}$, 6795 (2002). See Section V.

12. See Figs. 2 and 4 of F.J. Albert, N.C. Emley, E.B. Myers, D.C. Ralph and R.A. Buhrman, Phys. Rev. Lett. $\underline{89}$, 226802 (2002).

13. F. Schreiber, J. Pflaum, Z. Frait, Th. Muhge and J. Pelzl, Solid State Comm. $\underline{93}$, 965 (1995).

14. J.Y. Gu, S.D. Steenwyk, A.C. Reilly, W. Park, R. Loloee, J. Bass and W.P. Pratt, J. Appl. Phys. $\underline{87}$, 4831 (2000).

15. See Fig. 3 of L. Berger, J. Appl. Phys. $\underline{89}$, 5521 (2001).

16. S. Urazhdin, Norman O. Birge, W.P. Pratt and J. Bass, arXiv:cond-mat/0309191.

\vspace{1ex}

\hspace{4em} FIGURE CAPTIONS

\vspace{1ex}

FIG. 1. a) Multilayer for current-driven experiments, with current leads of same cross section area as the multilayer itself, where $F_{1}$ is the fixed magnetic layer and $F_{2}$ the free magnetic layer. b) Current-induced translation of spin-up and spin-down Fermi surfaces in momentum space. c) Current-induced expansion of spin-down and contraction of spin-up Fermi surfaces, near an interface.

FIG. 2. a) Multilayer for current-driven experiments, with current leads of large cross section in the shape of flat electrodes. Pillar length and layer thicknesses are smaller than shown here. b) Normalized spin accumulation $\overline{\Delta \mu}$ and $\Delta i$ versus normalized radial coordinate r, in the flat electrode of (a).

FIG. 3. Normalized spin accumulation $\overline{\Delta \mu}$ versus normalized coordinate x for the multilayer of Fig. 2a (curve labelled ``variable cross section'') and for that of Fig. 1a (curve labelled ``constant cross section''). For plotting purpose, the current density in the pillar is assumed to be $1A/m^{2}$. The parallel state is assumed and layer thicknesses $L_{F1}=10nm, L_{F2}=2.5nm,L_{N2}=6nm$.

FIG. 4. Normalized spin accumulation $\overline{\Delta \mu}$ in $F_{2}$ at x=0, versus thickness $L_{F1}$ of the fixed magnetic layer. Curves for parallel and antiparallel states are labelled P and AP, respectively. Another label indicates the nature of the current leads. For plotting purpose, the current density in the pillar is assumed to be $1A/m^{2}$. The thickness of the free magnetic layer and of the spacer are $L_{F2}=2.5nm,L_{N2}=6nm$.

FIG. 5. a) Ratio of critical current densities for parallel and antiparallel states versus thickness $L_{F1}$ of fixed magnetic layer, with leads of variable cross section. The thickness of the free magnetic layer and of the spacer are $L_{F2}=2.5nm,L_{N2}=6nm$. Open circles indicate experimental values of the ratio from various authors in Ref. 8. b) Magnetic poles on the surface of layer $F_{1}$ create a dipolar field $H^{D}_{z}$ at the $N_{2}/F_{2}$ interface. 

FIG. 6. Normalized values of critical current densities $j^{P}_{c},j^{AP}_{c}$ and of $j^{P}_{c}-j^{AP}_{c}$ versus thickness $L_{F1}$ of fixed magnetic layer, with leads of variable cross section. The thickness of the free magnetic layer and of the spacer are $L_{F2}=2.5nm,L_{N2}=6nm$. The dashed line represents the best fit of the theory to experimental values of $j^{P}_{c}-j^{AP}_{c}$ obtained by Albert et al. $^{12}$, and corresponds to $\alpha_{b}/\alpha_{s}=5.04$.

FIG. 7. Critical current densities $j^{P}_{c},j^{AP}_{c}$ normalized to their values for $L_{N2}=6nm$, plotted versus thickness $L_{N2}$ of spacer $N_{2}$, with leads of variable cross section. The two dashed curves represent the predictions for the P and AP state. The thicknesses of the fixed and free magnetic layers are $L_{F1}=10nm,L_{F2}=2.5nm$, respectively. The solid and open squares represent experimental values of $j^{P}_{c}$ and $j^{AP}_{c}$, respectively, obtained by Albert et al.$^{12}$ for three different $L_{N2}$ values.

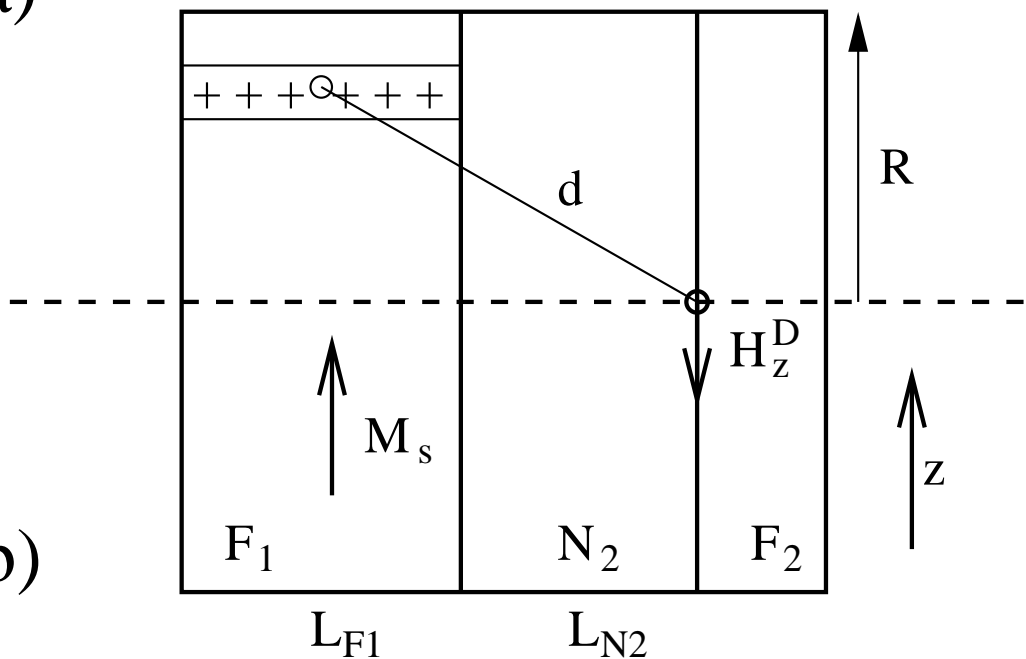
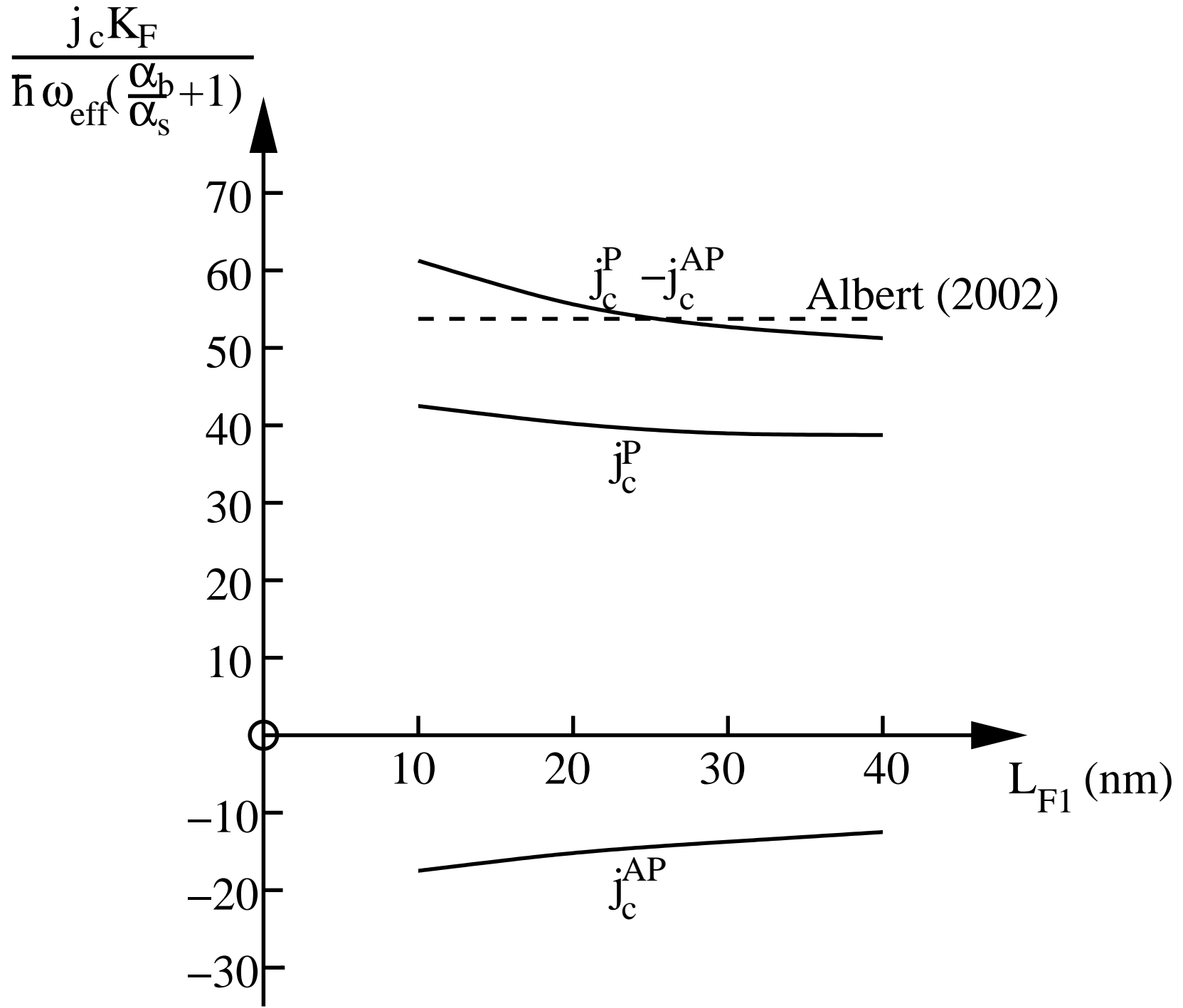
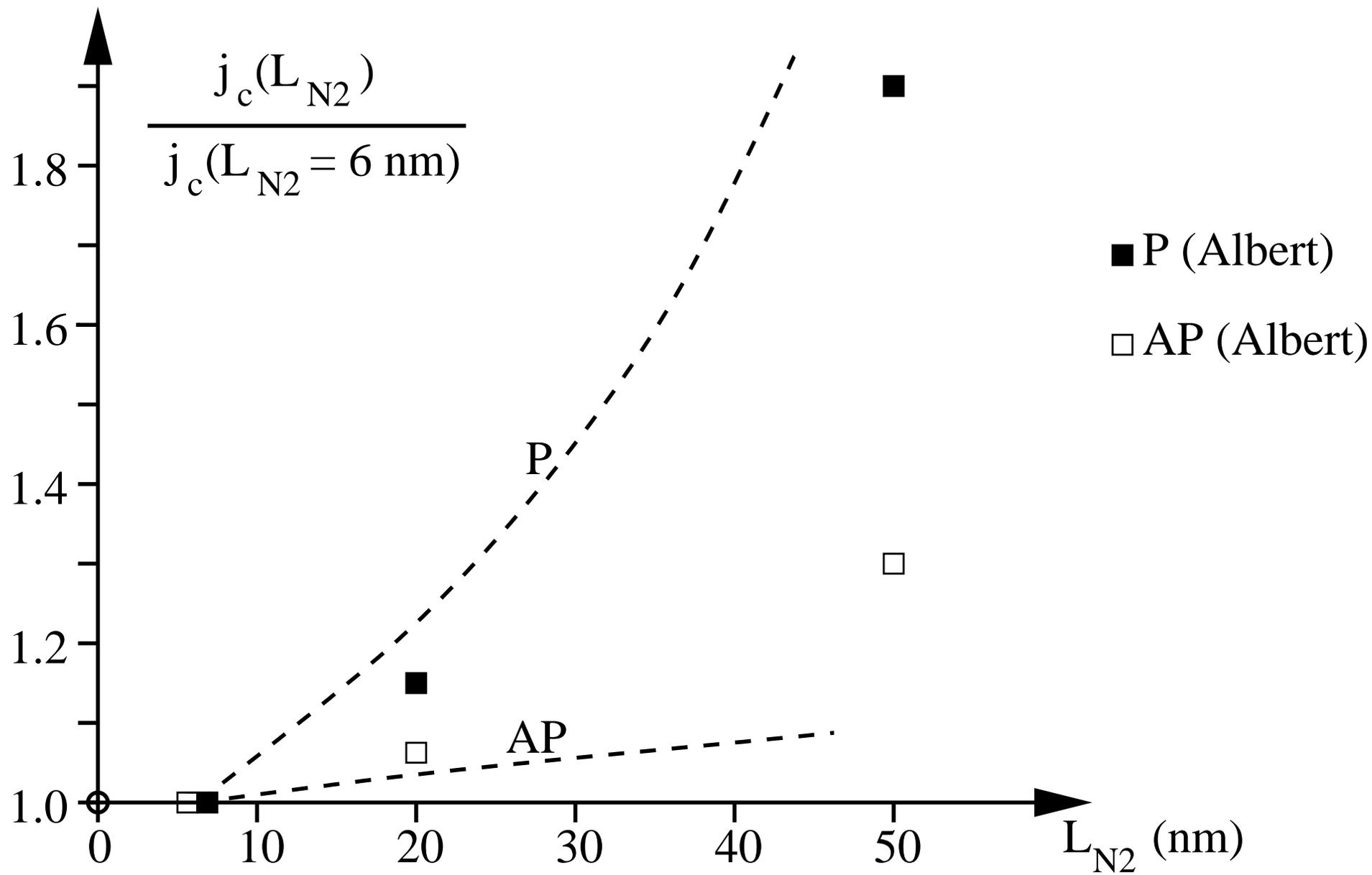
\end{document}